\documentclass[
	a4paper, % Paper size, use either a4paper or letterpaper
	10pt, % Default font size, can also use 11pt or 12pt, although this is not recommended
	unnumberedsections, % Comment to enable section numbering
	twoside, % Two side traditional mode where headers and footers change between odd and even pages, comment this option to make them fixed
]{LTJournalArticle}

\runninghead{WebRISC-V} % A shortened article title to appear in the running head, leave this command empty for no running head
\footertext{\small\textit{RISC-V Summit Europe, Paris 12-15th May 2025}}

\fancypagestyle{firstpagefooter}{%
    \fancyhf{}  % Clear all headers and footers
    \fancyfoot[RO]{\small\textbf{\thepage}}  % Page number on the right for first page
    \fancyfoot[LO,RE]{\footertext}  % Footer text
}

\title{WebRISC-V: A 64-bit RISC-V Pipeline Simulator \\ for Computer Architecture Classes} % Article title, use manual lines breaks (\\) to beautify the layout

\author{%
	Roberto Giorgi\textsuperscript{1,2} and Gianfranco Mariotti\textsuperscript{1}
    \thanks{Corresponding author: \href{giorgi@unisi.it}{giorgi@unisi.it}.
    This work is partly funded by the 
    Barcelona Zettascale Laboratory, promoted by the Spanish Ministry for Digital Transformation and the Civil Service, within the framework of the Recovery, Transformation and Resilience Plan - Funded by the European Union - NextGenerationEU
     and via the PNRR M4C2-Inv1.4 Italian Research Center on High-Performance Computing, Big-Data and Quantum Computing, cascade funding project EDGE-ME, MUR-ID: CN0000013.}%
}

\date{\footnotesize\textsuperscript{\textbf{1}}Department of Information Engineering and Mathematics, University of Siena, Via Roma 56, 53100 Siena, Italy\\ \textsuperscript{\textbf{2}}Barcelona Supercomputing Center (BSC), Barcelona, Spain}

\renewcommand{\maketitlehookd}{%
	\begin{abstract}
		\noindent WebRISC-V is a web-based educational tool designed to simulate the pipelined execution of assembly programs according to the RV64IM specifications (64-bit RISC-V processor). The tool allows users to investigate pipeline stalls, understand the internal state of pipeline architectural blocks, and visualize the cycle-by-cycle execution of instructions. WebRISC-V executes directly in a web browser, providing a detailed pipeline execution for RISC-V processors. This paper describes the features of WebRISC-V, compares it with similar tools, and provides an example of its usage in investigating the pipeline.
	\end{abstract}
}

\begin{document}
%

% SINGLE-COLUMN FIGURE
\newcommand{\MyFigure}[2][1.00]{% Number
\begin{figure}
\centering
\includegraphics[width=#1\columnwidth]{\MyPath/\MyPaper-figsm_FIG_#2.pdf}
\caption[]{\small \input{\MyPath/\MyPaper-figsm_Slide_#2.txt}}
\label{mypaper-fig#2}
\end{figure}}

% TWO-COLUMN FIGURE
\newcommand{\MyExtFig}[2][1.00]{% Number
\begin{figure*}[!htbp]
\centering
\includegraphics[width=#1\textwidth]{\MyPath/\MyPaper-figsm_FIG_#2.pdf}
%\caption[]{\footnotesize \input{\MyPath/\MyPaper-figs_Slide_#2.txt}}
\caption[]{\small\input{\MyPath/\MyPaper-figsm_Slide_#2.txt}}
\label{mypaper-fig#2}
\end{figure*}}

% FIGURE CITATION
\newcommand{\MyFigRef}[1]{% Number
Fig.~\ref{mypaper-fig#1}}

\maketitle % Output the title section
\thispagestyle{firstpagefooter} % Apply first page style
\pagestyle{fancy}  % Apply standard fancyhdr style for the rest
%----------------------------------------------------------------------------------------
%	ARTICLE CONTENTS
%----------------------------------------------------------------------------------------
\section{Introduction}
Instruction pipelining is a fundamental concept in Computer Architecture courses, as it significantly improves processor performance. However, its real impact is often underestimated or misunderstood by students. WebRISC-V addresses this educational gap by offering an interactive, web-based tool that allows users to visualize and analyze pipeline execution in a user-friendly manner. 

Understanding pipeline behavior, including stalls and hazards, is crucial in optimizing processor performance. WebRISC-V provides a cycle-by-cycle analysis of RISC-V instructions in a pipeline, allowing users to gain a deeper understanding of execution flow, instruction dependencies, and bottlenecks that affect speedup.

\section{Key Features and Contributions}
WebRISC-V introduces several innovations that distinguish it from existing tools, including:
\begin{itemize}
    \item A browser-based interface requiring no installation, making it accessible from any device.
    \item Cycle-accurate visualization of pipeline execution, via instruction flow and hazard detection.
    \item Identification and classification of stalls, hazards, and structural dependencies within the pipeline.
    \item Interactive features for modifying instructions and observing real-time execution changes.
    \item A side-by-side comparison of different execution sequences, providing insights into pipeline optimization.
    \item Automatic generation of detailed pipeline execution diagrams, aiding in teaching and research.
    \item A comparison with other available pipeline visualization tools, highlighting WebRISC-V’s advantages in usability and accuracy.
\end{itemize}

\section{Other Existing Pipeline Tools}
Several tools exist for visualizing pipeline execution, but most require local installation and lack detailed cycle-by-cycle analysis. WebRISC-V stands out due to its web-based accessibility and real-time feedback. Tools such as Ripes and QtSPIM provide some pipeline visualization but do not offer the level of detail or interactivity as WebRISC-V. By allowing users to interact directly with the execution process, WebRISC-V serves as a more effective learning tool.
PBSE,
%\cite{Lim19}
(MARS plug-in),
%\cite{Vollmar06}
MIPS X-Ray
%\cite{Sales10}
(MARS plug-in),
DrMIPS,
%\cite{Nova13}, 
Mipster32,
%\cite{DeOliveiraQuintas16}, 
UCOMIPSIM,
%\cite{Gersnoviez18}, 
Visimips,
%\cite{Kabir11}, 
WASP 
%\cite{Stojkovic07} 
and WebMIPS
%\cite{Branovic04} 
are MIPS ISA tools; 
Ripes \cite{Ripes} and WebRISC-V are similar tools, but supporting the RISC-V ISA.
While Ripes supports gcc-compiled code and is more tailored to developers, WebRISC-V restricts the supported ISA to `RV64IM' to provide an environment more focused on educational principles.

\subsection{Software Architecture}
WebRISC-V has its back-end written in PHP and its front-end in HTML, CSS and JavaScript \cite{Nixon12}.
Being a server-side web application, it is installed and executed on a web server and presented to the user on their client interface.
If the teaching staff wants a local installation, an instance can be hosted with a simple procedure on a Linux or Windows server.

This simulator supports the full implementation of two RISC-V `modules'\footnote{In a few words, a RISC-V `module' is a subset of instructions.} as they are described in the RISC-V ISA unprivileged specification \cite{spec}: the 64-bit Base Integer (``fence'' instruction excluded) module - also called `RV64I' module - and the Extension for Integer Multiplication and Division module - also called `M' module, therefore making the tool run according to the `RV64IM' specification.

Voluntarily, for easier student reference, WebRISC-V closely resembles the schematic 
used in the Patterson/Hennessy book \cite{Patterson17},
in which the pipelined datapath implementation is explored and explained.

%that appears in
%the book ``Computer Organization and Design RISC-V Edition'' by D. A. Patterson and J. L. Hennessy \cite{Patterson17},
%in which the pipelined datapath implementation is explored and explained.
\begin{figure} [!hb]
	\centering
	\includegraphics[width=\columnwidth]{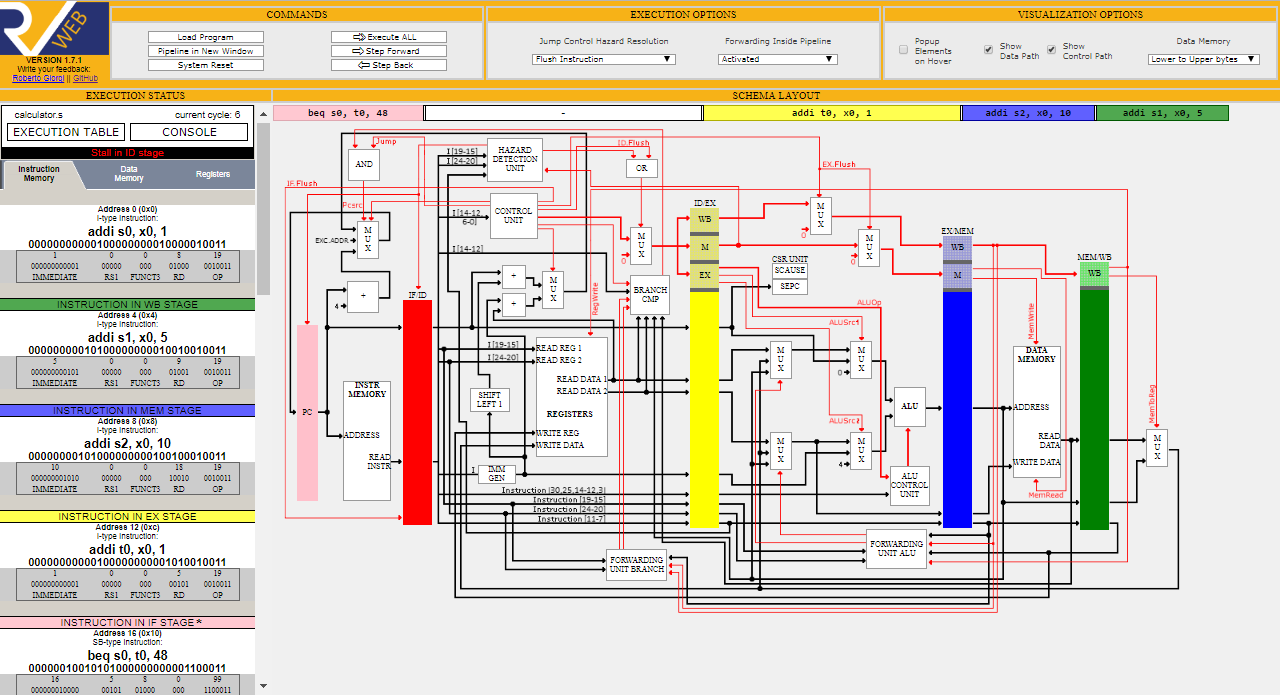}
	\caption{
		WebRISC-V main page.
	}
	\label{intro}
\end{figure}

\subsection{Software Functionalities of WebRISC-V}
\begin{figure}
	\centering
	\includegraphics[width=\columnwidth]{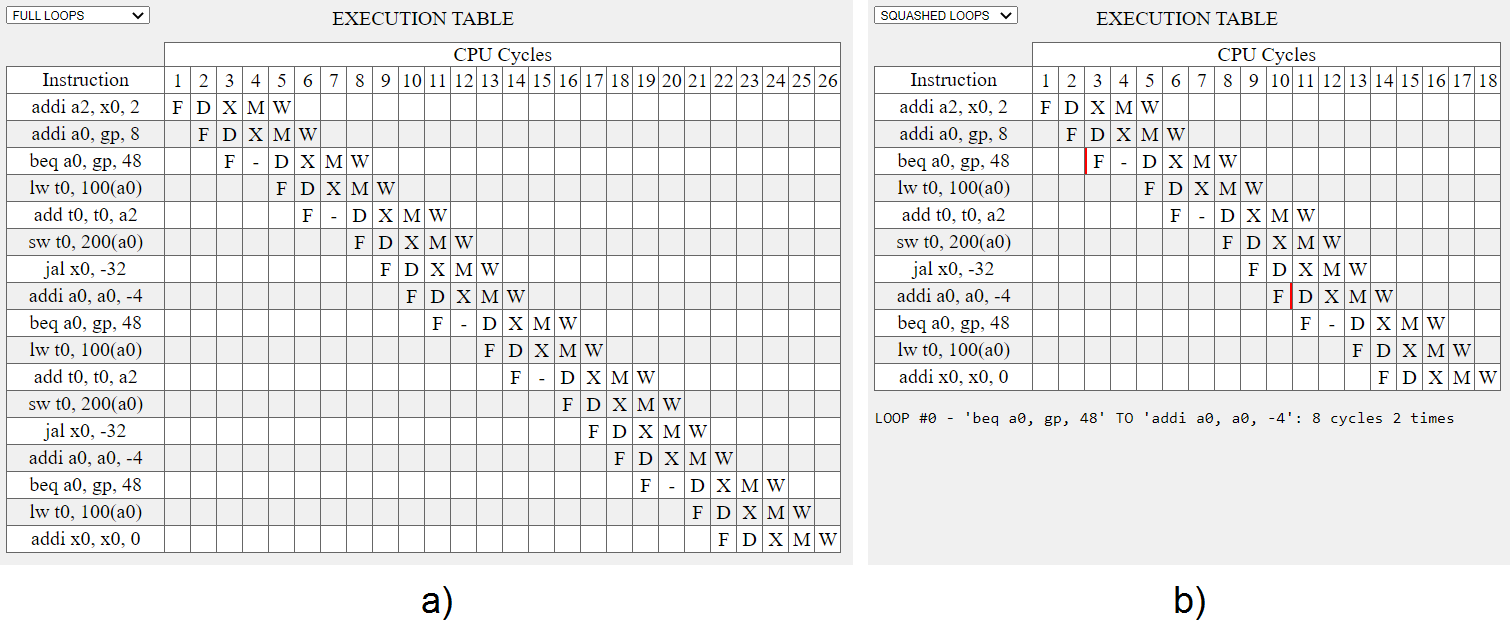}
	\caption{
		Pipeline diagram in both visualization methods: (a) `Full loops' and (b) `squashed loops'.
	}
	\label{exc_tbl}
\end{figure}

\begin{itemize}
  \item The visualization of the complete architectural schematic of a pipelined RISC-V  (see Fig.~\ref{intro}).
  \item The ability to step forward and backward in the execution of code, to better study what is happening inside the pipeline and its elements.
  \item On a single page view, the monitoring of the information about the current processing state (e.g., cycle count, colored tags to indicate the current stage of an instruction, highlighting of eventual 'bubbles' in the pipelined execution).
  \item A descriptive explanation of each internal element together with its current state, that can be shown by simply hovering with the mouse.% (see Fig.~\ref{stall_hzd_unit});
  \item Forwarding/No-Forwarding simulation modes.
  \item The possibility of enabling
  %(see Fig.~\ref{main_w_wo_fwd}a)
  or disabling
  %(see Fig.~\ref{main_w_wo_fwd}b) 
  the data forwarding units (with automatic visualization of the corresponding schematic).
  \item The possibility of visualizing the memory segment contents (Text Segment, Static Data Segment, Dynamic Data Segment) and the registers. %, as shown in Fig.~\ref{mem_reg};
  \item An online editor, with some built-in examples, and a contextually visible full list of the available instructions and directives. % (see Fig.~\ref{editor});
  \item Automatic generation of the classic pipeline diagram
  (see Fig.~\ref{exc_tbl}a); 
  in case of loops, this diagram can be automatically squashed
  (see Fig.~\ref{exc_tbl}b).
  \item Basic I/O system calls of the simulated RISC-V are possible by prompting the user via a popup window that emulates the system console.
\end{itemize}

\section{Use Cases in Education}
WebRISC-V is designed for educational use, particularly in undergraduate computer architecture courses. Professors can use it to demonstrate pipeline execution principles, while students can experiment with different instruction sequences to observe their effects on performance. The tool aids in understanding the impact of forwarding, stalls, and branch hazards, making abstract concepts more tangible. Moreover, its web-based nature ensures accessibility without installation barriers, facilitating both classroom and remote learning \cite{Mariotti22-softwarex}.

\subsection{GitHub Repository}
Source code and documentation are publicly available: %\newline
\url{https://github.com/Mariotti94/WebRISC-V}

%\section{Conclusion and Future Work}
%WebRISC-V enhances the understanding of instruction pipelining by offering an accessible, interactive, and detailed visualization of RISC-V execution. Its web-based nature eliminates software installation barriers, making it ideal for educational use. Future work includes extending its support to additional RISC-V features, integrating multi-core simulation, and improving visualization capabilities. Additional research will explore how WebRISC-V can aid students in mastering advanced computer architecture topics, such as out-of-order execution and speculative execution.

%\section*{Acknowledgements}
%This work is partly funded by Barcelona Zettascale Laboratory, promoted by the Spanish Ministry for Digital Transformation and the Civil Service, within the framework of the Recovery, Transformation and Resilience Plan - Funded by the European Union - NextGenerationEU and via the PNRR M4C2-Inv1.4 Italian Research Center on High-Performance Computing, Big-Data and Quantum Computing, cascade funding project EDGE-ME, MUR-ID: CN0000013.
%----------------------------------------------------------------------------------------
%	 REFERENCES
%----------------------------------------------------------------------------------------

\printbibliography % Output the bibliography

%----------------------------------------------------------------------------------------

\end{document}